\documentclass[mathpazo]{cicp}

\usepackage{bm}
\usepackage{tikz}

\newcommand\diff{\,\mathrm d}
\newcommand\mi{\mathrm i}
\def\revisedcolor{black}

\begin{document}
\title{Modeling of bound electron effects in particle-in-cell simulation}



 \author[An X Y et.~al.]{Xiangyan An\affil{1}\affil{2}, Min Chen\affil{1}\affil{2}\comma\corrauth,
       Zheng-Ming Sheng\affil{1}\affil{2}\affil{3}, and Jie Zhang\affil{1}\affil{2}\affil{3}}
 \address{\affilnum{1}\ Key Laboratory for Laser Plasmas (MoE), School of Physics and Astronomy, Shanghai Jiao Tong University, Shanghai 200240, China \\
           \affilnum{2}\ Collaborative Innovation Center of IFSA, Shanghai Jiao Tong University, Shanghai 200240, China\\
           \affilnum{3}\ Tsung-Dao Lee Institute, Shanghai Jiao Tong University, Shanghai 200240, China}
\emails{{\tt minchen@sjtu.edu.cn} (M.~Chen)}

\begin{abstract}
To include the bound electron effects in particle-in-cell (PIC) simulation, we propose a model in which the response of the dipole components of partially ionized ions to external electromagnetic fields can be included.  Instead of treating the macro-ion particle as a single particle without an internal structure, the ions are considered to have a structure composed of a central nucleus and a bounded electron cloud in our model. The two parts experience the interactions of both the external electromagnetic fields and the internal Coulomb fields. In this way, the laser scattering effects by a partially ionized medium can be modeled properly in the PIC simulation. The laser propagation in a neutral medium and the Bragg scattering of the laser in crystal structure have been simulated with a PIC code modified based on our model as the benchmark. Our model may find applications to study some interesting problems, such as the x-ray laser-driven wakefield acceleration in crystals, the x-ray laser-driven high energy density physics, and intense laser propagation in partially ionized nonlinear optical materials, etc.
\end{abstract}


\maketitle

\section{Introduction}
\label{sec-introduction}

Particle-in-cell (PIC) codes are widely used in plasma physics, especially in the simulation of laser-plasma interaction\cite{dawson_particle_1983}, such as in inertial confinement fusion\cite{nishimura_energy_2011}, laser wakefield acceleration\cite{tajima_laser_1979,esarey_physics_2009}, high order harmonics generation\cite{teubner_high-order_2009} and so on. However, the usual PIC codes only consider the electric currents contributed by free electrons and ions. These charged particles are modeled by particle clouds with special shapes and finite size, \textit{i.e.} macro-particles. They are accelerated by external electromagnetic (EM) fields which are calculated from the Maxwell equations with source terms contributed by the charged particles. The internal EM properties of some particles, such as the dipole components contributed by the bound electrons and the central nucleus, are neglected. For particles with the same charge to mass ratio, their trajectories would be identical once their initial positions and velocities are the same. For example, the dynamical evolutions of a proton and a He$^+$ ion in normal PIC simulation are the same although there is one bound electron in the He$^+$ ion. Even though the ionization process is included in some PIC codes \cite{arber_contemporary_2015,chen_numerical_2013}, the effects of the bound electrons of the ions are still neglected, where only the free electrons which are ionized will respond to the external EM fields. 

On the other hand, in some applications, where the laser fields are not high enough to ionize all the electrons of the ions and the electromagnetic response of the bound electrons cannot be neglected, the usual PIC codes will be inadequate.  In the wakefield acceleration by x-ray pulses in crystal~\cite{tajima_crystal_1987}, nanotubes, and dense plasmas~\cite{shin_x-ray_2013,zhang_particle--cell_2016,hakimi_x-ray_2019,tajima2020}, for example, although the x-ray propagation and the following wakefield acceleration have been numerically studied and even the radiation reaction and the collision have also been included\cite{dodin_charged_2008}, the response of the internal bound electrons has not been included. Some researchers have placed the ions periodically in space to simulate the crystals effect~\cite{hakimi_wakefield_2018} and the fine modulation of the wake was observed in the simulations~\cite{liang_high-repetition-rate_2020}. However, the laser diffraction by the crystal structure cannot be properly simulated since this effect is related to the response of the internal bound electrons.

In this paper, we extend the usual PIC algorithm to include the effects of the bound electrons. The ions that are not fully ionized are considered a two-body macro-particle. The Coulomb interaction connects the two bodies. Since one macro-particle in the PIC codes represents many real particles with similar trajectories in the phase space. To extend this model to the bound electrons, some limitations should be satisfied for specific applications. Otherwise, the effect of the bound electrons may be amplified due to the artificial coherence induced by the macro-particle. As long as the size of the macro-particles ($\lambda_{pseu}$, usually characterized by the grid resolution $\diff x$) is much smaller than the character scale length of the concerned problem ($\lambda_c$, usually characterized by the laser wavelength $\lambda_0$, radiation wavelength or the plasma wavelength), the macro-particle can be used not only to represent the trajectory of the represented particles; but also to correctly handle the coherent synthesis effects of these particles. In this case, although the electrons are combined to different ions, their responses to the external EM fields can be coherent since $\lambda_{pseu}<\lambda_c$. The artificial coherence embedded in the macro-particle model correctly represents the real physical effect. Another issue is the different responses of each bound electron in the same ion. Usually, the bound electrons are in different quantum states. The outer ones may show a stronger response to the external fields since the Coulombic force from the ion is relatively weak. However, it is impossible to include these detailed quantum responses of different bound electrons in a PIC code and it is not necessary for many applications. In our macro-particle model, a combined electron in the macro-particle is represented by $Z_b$ bound electrons of the ions and $Z_b$ is an adjustable number determined by the ionization degree. The detailed responses of every single electron are neglected, only the main response contributed by the dipole component has been reserved. That is to say that the code is only useful when a localized collective electron in an ion is a useful approximation. It cannot be used in the processes where the quantum effects, such as excitation and ionization by x-ray, dominate. These processes usually need to include atomic physics in PIC codes by storing the internal states and calculating the transition matrix. The quite different spatial and temporal scales for atomic process and plasma evolution should be involved, which is almost impossible for today's computational technology. However, for our targeted problems, the plasma effects dominate over the quantum response of the ions.

\textcolor{\revisedcolor}{We also noticed x-ray plasma interaction has been modeled in some existed code \cite{chung_atomic_2017,royle_kinetic_2017}, where x-ray laser is modeled by photons and its interaction with plasmas is described by the transport equation $\left(\frac{1}{c}\frac{\partial}{\partial t} +\bm n\cdot\bm{\nabla}\right)I=\eta-\chi I $, where $I(\bm r,\Omega,\nu,t)$ is the x-ray intensity, $\eta(\bm r,\nu,t)$ is the emissivity, $\chi(\bm r,\nu,t)$ is the opacity, $\Omega(\theta,\phi)$ is a solid angle, and $\nu$ is the radiation frequency. In this approach, the detailed responses of the ions to the x-ray are represented by the plasma's opacity. Although the x-ray transportation can be modeled, the evolution of the electromagnetic fields with the x-ray frequency cannot be explicitly obtained, which is important for the motion of the freely charged particles. Differently from that, in our model, the x-ray laser is treated as EM waves, and the response of the bound electrons is calculated according to the particle model.} In our way, both the EM response of the neutral medium and the Bragg diffraction effect of laser propagation in the crystal can be modeled properly. In the following, we will introduce the modified particle model and show some typical simulation cases for the benchmark of our model.

\section{Two-Body Macro-Particle Model for Ions}
\label{sec-model}
The ions and electrons in the usual PIC codes are represented by macro-particles. The weight of a macro-particle means the number of the real particles in it. The motions of these macro-particles with mass $m_c$, charge $q_c$ and velocity $\bm v_c$ are calculated according to the external EM fields. The term $q_c\bm v_c/\diff V$ is included in the current to update the fields.
\begin{figure}[ht]
			\centering
			\includegraphics[scale=1.4]{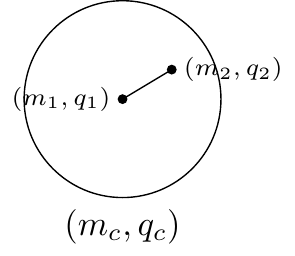}
			\caption{The charge and mass distribution of the macro-particle for ions that are not fully ionized in our extended PIC code.\label{fig-ion}}
\end{figure}

As shown in Fig.\,\ref{fig-ion}, in our extended PIC simulation model, we assume an partially ionized ion particles is composed of two parts: a nucleus with mass $m_1$, charge $q_1$, and a combined bound electron with mass $m_2$, charge $q_2$. The particle also contains a bound electron number $Z_b$. Thus $m_{1,2}$ and $q_{1,2}$ can be determined according to $q_2 = -Z_be, m_2=Z_bm_e, q_1 = q_c-q_2, m_1=m_c-m_2$, where $e$ and $m_e$ represent the charge and mass of an electron. Either of them has its own position and momentum. At each time step, we update their momenta according to the Boris algorithm \cite{birdsall_plasma_nodate}.
\begin{gather}
	\bm u^-_{1,2}=\bm u^n_{1,2}+\frac{q_{1,2}\Delta t}{2m_{1,2}}\bm E_{1,2}(\bm r^{n+1/2}_{1,2}) \\
		\bm u^+_{1,2}=\bm u^-_{1,2}+\left(\bm u^-_{1,2}+\left(\bm u^-_{1,2}\times\bm t_{1,2}\right)\right)\times\bm s_{1,2} \\
		\bm u^{n+1}_{1,2} = \bm u^+_{1,2}+\frac{q_{1,2}\Delta t}{2m_{1,2}}\bm E_{1,2}(\bm r^{n+1/2}_{1,2}) \\
		\text{with} \quad\bm t_{1,2} = \bm B_{1,2}(\bm r^{n+1/2}_{1,2})q_{1,2}\Delta t/(2m_{1,2}\gamma^-_{1,2}),\quad \bm s_{1,2}=2\bm t_{1,2}/(1+\bm t_{1,2}^2)
\end{gather}
where $\bm u, q, m, \Delta t, \bm E, \bm B, \bm r$ are the nomalized momentum $p/(mc)$, the charge, the mass, the time step, the electric field, the magnetic field, and the position, respectively, and the subscript $1,2$ represents the quantities of the nucleus and the electron, respectively. The electric fields here contain not only the external fields, but also the internal interaction from the nucleus and the combined electrons which can be calculated from the following modified Coulomb potential. Then the currents of each part $q_{1,2}v_{1,2}/\diff V$ are included into the total current, which is used to update the fields.

Since we introduce new features in PIC simulation, their initial conditions should be set. At this point, by considering the momentum conservation and assuming the fact that there are no electromagnetic fields emitted from the ions at the beginning, one can get:
\begin{gather}
   \bm p_1 + \bm p_2 = \bm p_c, \\
	q_1\bm v_1+q_2\bm v_2 = 0.
\end{gather}
The total momentum $\bm p_c$ depends on the initial temperature of the atoms or the partially ionized ions. If we set it as zero, the corresponding initial condition for the nucleus and the combined electron is $\bm v_1=\bm v_2 =0$.

\begin{figure}[ht]
	\centering
	\includegraphics[scale=0.5]{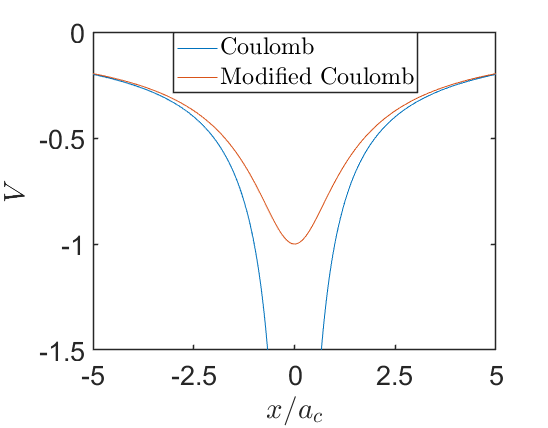}
	\caption{The comparison of usual Coulomb interaction $V=-1/r$ and the modified one in our PIC code $V=-1/\sqrt{r^2+a_c^2}$.\label{fig-coulomb}}
\end{figure}

Since the main response is the average dipole contribution of the combined electrons, to exclude the extreme response from the innermost electrons, we use a modified Coulomb potential, i.e. $V_M=\frac{q_1q_2}{4\pi\varepsilon_0 \sqrt{r^2+a_c^2}}$ to simulate the interaction between the nucleus and the combined electrons at the out shells~\cite{hu_phase_2004,hu_intense_2004}, where $r$ represents the distance between the combined electron and nucleus. It corresponds to a modified Coulomb force $\frac{q_1q_2\bm r}{4\pi\varepsilon_0 (r^2+a_c^2)^{3/2}}$. Here $a_c$ is an adjustable parameter determined by the material. The numerical divergence of the usual Coulomb force with $r=0$ is also avoided. The comparison of the Coulomb interaction and the modified one is shown in Fig.\,\ref{fig-coulomb}. As one can see, the field is close to the usual Coulomb field for the out shell electrons.

Since the modified Coulomb interaction vanishes when the nucleus and the electron are at the same position, the initial condition $\bm r_1=\bm r_2=\bm r_c$ should be taken, where $\bm r_c$ is the position of the ion.

\section{Benchmark of the new PIC model}
\label{sec-simulation}
Based on the above model, we have modified the PIC code EPOCH\cite{arber_contemporary_2015} to include the effects of the combined electrons. \textcolor{\revisedcolor}{The positions and the momenta of the nucleus and the bound electrons are added as new data elements of the ion class and the whole structure of the code is thus maintained. We just need to update the new features of the ions in the usual calculation loop and the data transfer among the different parallel processes is almost the same as before. Thus the performance of the code is not affected a lot.}

To check the validity of our modified code, we apply it to two conditions. First, we study the laser propagation in a neutral medium where all the particles are neutral without plasma. We theoretically deduce the relationship between the material's permittivity and the parameters of the macro-particles used in the code. Simulations are then performed to check its validity. Then we study the laser propagation in a crystal, where the ions are periodically placed and the Bragg effect can be observed when the ions are not fully ionized.

\textcolor{\revisedcolor}{Since one of the main purposes of our code is in the study of x-ray laser-driven wakefield acceleration, in the following, we use x-ray as the drive laser. It is quite natural to extend such applications to the usual laser wavelength such as the Ti Sapphire laser with a wavelength of 800nm. }

\subsection{Simulation of x-ray laser propagation in neutral medium}
In the usual PIC simulations, the vacuum permittivity is used in the Maxwell equations. No material permittivity is included because the material polarization contributed by the combined electrons is not considered. In our extended PIC code, such contribution and the permittivity of the material can be naturally included. Since we have used the new macro-particle model, the vacuum permittivity is still used in the Maxwell equation. However, we should get the relation between the macro-particle model and the material permittivity.

For simplicity, we take the case of non-relativistic laser intensity so that the magnetic field of the laser and the longitudinal motion of the particles can be ignored. Then one can obtain the equations of motion for the nucleus and the electron.
\begin{gather}
		m_1\frac{\diff^2 r_1}{\diff t^2}=q_1E+F(r), \\
		m_2\frac{\diff^2 r_2}{\diff t^2}=q_2E-F(r), \\
\end{gather}
where $F(r)$ is the internal Coulomb force between the two bodies within the macro-particle of the ions. Since we have modified Coulomb potential to $V_M=\frac{q_1q_2}{4\pi\varepsilon_0 \sqrt{r^2+a_c^2}}$, the intrinsic frequency of the system can be obtained from the Taylor series of $V_M$:
\begin{gather}
	\frac{1}{2}\left.\frac{\partial^2 V}{\partial r^2}\right|_{r=0}=\frac{1}{2}m_r\omega_c^2 \\
		\omega_c^2=\frac{\left|q_1q_2\right|}{4\pi\varepsilon_0m_ra_c^3}.
\end{gather}

Then the equations of motion can be solved and we find
\begin{gather}
	r_1=\left[-\frac{q_c}{m_c\omega^2}+\frac{m_2}{m_1+m_2}\cdot\frac{q_r}{m_r(\omega_c^2-\omega^2)}\right]E, \\
	r_2=\left[-\frac{q_c}{m_c\omega^2}-\frac{m_1}{m_1+m_2}\cdot\frac{q_r}{m_r(\omega_c^2-\omega^2)}\right]E,
\end{gather}
where $E$ and $\omega$ are the electric field and angular frequency of the laser, respectively, and $(m_1+m_2)r_c=m_1r_1+m_2r_2,\,\,r=r_1-r_2,\,\,m_c=(m_1+m_2),\,\,q_c=(q_1+q_2),\,\,m_r=(m_1m_2)/(m_1+m_2),\,\,q_r=(q_1m_2-q_2m_1)/(m_1+m_2)$.

Then we can derive the permittivity of such a medium through $D=\varepsilon E=\varepsilon_0 E+P=\varepsilon_0 E+\sum\mathcal P/\diff V=\varepsilon_0+n_1q_1r_1+n_2q_2r_2$, where $\mathcal P$ is the dipole moment of such an atom and $P$ is the electric polarization. The result is
\begin{gather}
	\begin{split}
	\varepsilon=\varepsilon_0&+\left\{n_1q_1\left[-\frac{q_c}{m_c\omega^2}+\frac{m_2}{m_1+m_2}\cdot\frac{q_r}{m_r(\omega_c^2-\omega^2)}\right]\right. \\
			&\quad\,\,\,\left.+n_2q_2\left[-\frac{q_c}{m_c\omega^2}-\frac{m_1}{m_1+m_2}\cdot\frac{q_r}{m_r(\omega_c^2-\omega^2)}\right]\right\}.
	\end{split}
\end{gather}
This formula connects the macro-particle parameter $a_c,\,Z_b$ with the material's property $\varepsilon$ \textcolor{\revisedcolor}{ and the dispersion relationship in such a material is determined by $c^2k^2/\omega^2=\varepsilon/\varepsilon_0=\varepsilon_r$.}

\begin{figure}[htbp]
	\centering
	\includegraphics[scale=1.2]{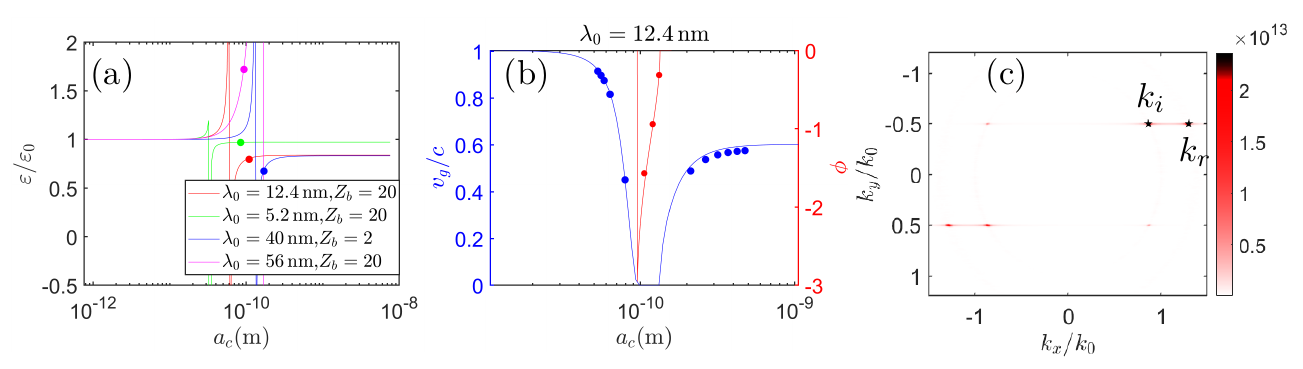}
	\caption{ \label{fig-neutralresult}(a) The permittivities calculated by our model with different macro-particle parameters (solid lines) and the data from the results of Hagemann\cite{Hagemann:75} (scattered points). (b) The group velocities of the x-ray laser in Au \textcolor{\revisedcolor}{and the phase-shifts in the reflection} calculated by the model with different macro-particle parameters (solid lines) and the results from the PIC simulations (scattered points).\textcolor{\revisedcolor}{(c) The Fourier transformation of the electric field when a laser beam is obliquely incident to the gold for the case $\lambda_0=56\,\text{nm},\,Z_b=20$ in (a). The pentagrams mark the wave vectors $k_i,\,k_r$ of the incident and refractive laser beam, respectively.}}
\end{figure}

To illustrate the results more clear, we take the material gold, for example, which is $n_2=n_1=n_0=5.9\times 10^{22}\,\text{cm}^{-3}$. 
Meanwhile, since the gold has one valent electron, we have considered these electrons as free, which will add a term $-n_0e^2/(m_e\omega^2)$ in the permittivity. As shown in Fig.\,\ref{fig-neutralresult}-(a), the permittivity in our model can be larger than one, which is impossible for the usual plasmas. The divergences come from the resonance when $\omega\sim\omega_c$. Meanwhile, we can always choose proper parameters $Z_b$ and $a_c$ to obtain the real permittivity.


We have carried out PIC simulations to check the model and the modified code. In the simulation, we use the same laser with the wavelength of 12.4\,nm(corresponding to photon energy 0.1\,keV) and the normalized vector potential $a=\left|eE/m_e\omega c\right|=1.0\times10^{-6}$. In addition, the gold atoms are set to be partially ionized with $Z_b=78$ and there is one free electron per Au atom in the simulation. We let the laser propagate in the medium for $300T_0$ with different $a_c$ and compare the group velocities obtained from the simulation and the model. \textcolor{\revisedcolor}{The group velocity in a material with permittivity $\varepsilon$ is given by
\begin{gather}
	v_g = \frac{\diff\omega}{\diff k}=\left(\frac{k}{\omega}+\frac{\omega^2}{2c^2k}\cdot\frac{\diff\varepsilon_r}{\diff\omega}\right)^{-1}
\end{gather}}
The result is given in Fig.\,\ref{fig-neutralresult}-(b), which shows that the group velocities from the simulations match well with those from the model. \textcolor{\revisedcolor}{The group velocity falls fast around $a_c\sim 9.5\times10^{-11}\,\text{m}$, which comes from the resonance around $\omega\sim\omega_c$. Around this region, the PIC simulations show that the laser is reflected and the group velocity is meaningless there. Thus we compare the phase-shifts $\phi$ in the reflection instead. The complex amplitude of the incoming and reflecting electric fields can be described by
\begin{gather}
		\tilde{E}_{0_R}=\left(\frac{1-\tilde{\beta}}{1+\tilde{\beta}}\right)\tilde{E}_{0_I} \\
		\tilde{\beta}=\sqrt{\frac{\varepsilon}{\varepsilon_0}}=\Gamma\mi
	\end{gather}
Then the phase-shift $\phi$ is given by
\begin{gather}
	\phi=\arctan\left(-\frac{2\Gamma}{1-\Gamma^2}\right)
\end{gather}
As one can see that the simulated phase shifts in these cases agree with the theoretical prediction quite well.} 

\textcolor{\revisedcolor}{Moreover, we also take the case $\lambda_0=56\,\text{nm},\,Z_b=20$ in Fig.\,\ref{fig-neutralresult}-(a) as a benchmark of refraction. The laser beam is set to be incident to the gold with an incident angle $\pi/6$. Then the transformation of the electric field of the incident and refractive laser is shown in Fig.\,\ref{fig-neutralresult}-(c). Again, our code gives a correct angle of refraction. These simulations show the validity of our code to describe the x-ray laser propagation in a neutral medium.
}



\subsection{Simulation of x-ray laser diffraction in a crystal}

\begin{figure}[htbp]
		\centering
		\includegraphics[scale=0.8]{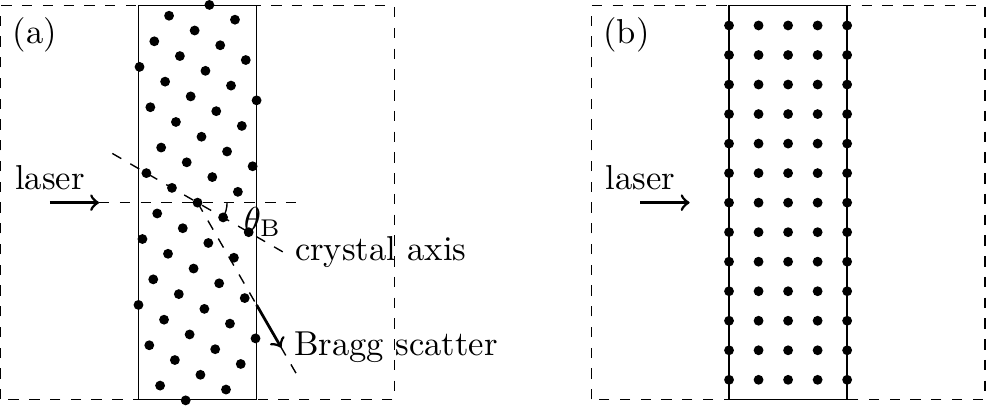}
		\caption{\label{fig-sketch}The sketch map of the simulations of Bragg effect.\label{fig-sketch}(a) The crystal axis is set to be in Bragg angle to the $\bm x$ axis. (b)The crystal axis is set to be in the direction of the $\bm x$ axis.}
	\end{figure}
	
	\begin{figure}[htbp]
		\centering
		\includegraphics[scale=0.7]{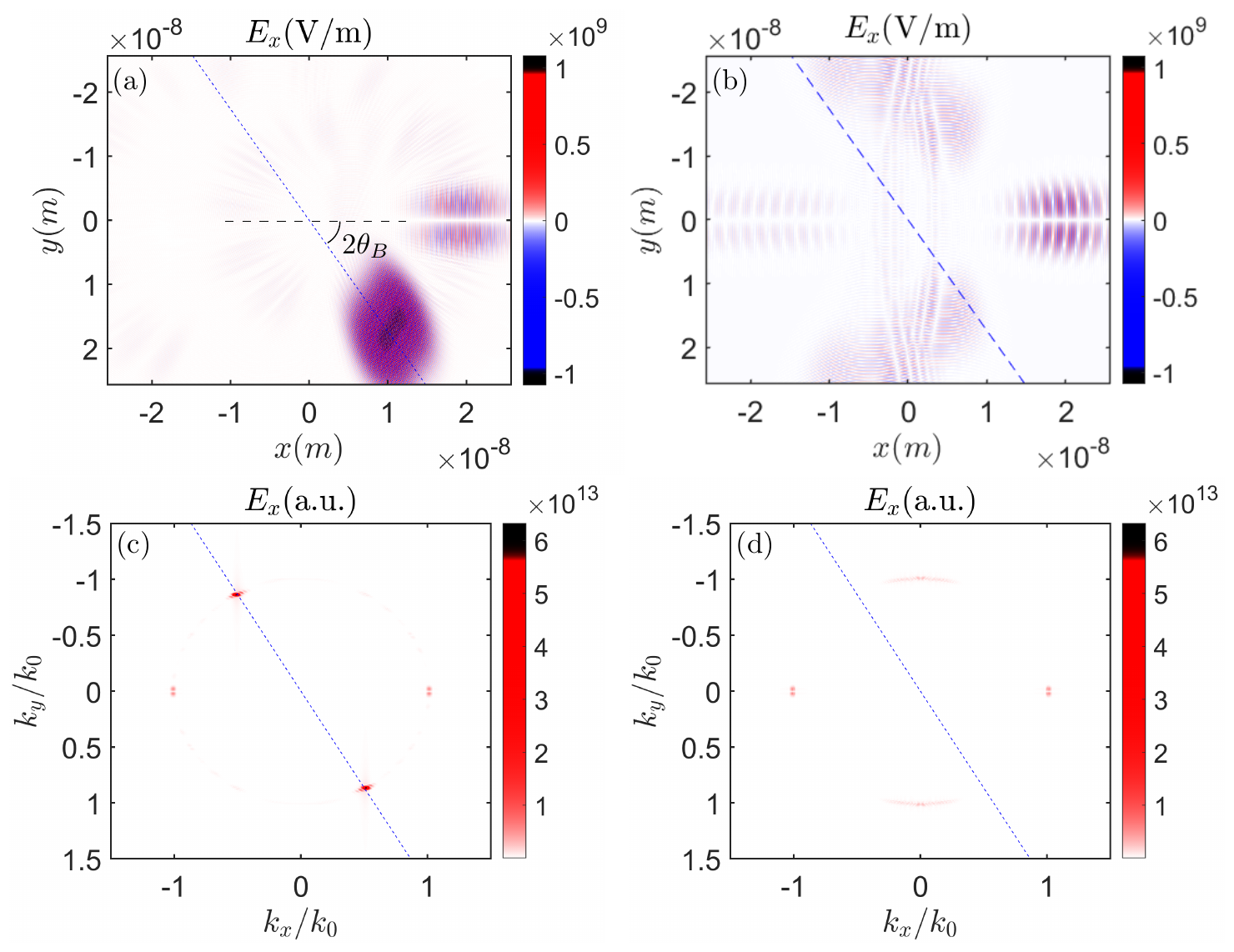}
		\caption{\label{fig-weak}The $E_x$ fields from the simulations after the laser transmits the crystal. The dashed line shows the direction of Bragg scattering. 
		 The $E_x$ field (a) and the 2-D Fourier transformation (c) when there is a Bragg angle between the laser and the crystal axis. The $E_x$ field (b) and the 2-D Fourier transformation (d) when there is no angle between the laser and the crystal axis.}
	\end{figure}

As mentioned in the introduction, to correctly simulate the wakefield acceleration driven by x-ray lasers in crystals, the laser propagation in a not fully ionized periodic structure should be carried out. In such cases, the Bragg diffraction is a typical phenomenon. To simplify the problem, we only consider a two-dimensional crystal structure. The material is chosen to be gold and the number density of the neutral gold atoms is set to be $n=5.9\times10^{22}$\,cm$^{-3}$. Thus the crystal constant is set to be $d=n^{-1/3}=0.26\,\text{nm}$. The wavelength $\lambda_0$ of the x-ray laser is set to be equal to the crystal constant. The laser has a width of 3\,nm and a pulse length of 5\,nm. The peak electric field of the x-ray laser is set to be $1.23\times10^{10}$\,V/m. The simulation box is set to be $[-100\lambda_0,100\lambda_0]\times[-100\lambda_0,100\lambda_0]$ and the gold crystal covers the region $[-20\lambda_0,20\lambda_0]\times[-100\lambda_0,100\lambda_0]$. The laser propagates along the $\bm x$ axis. With these parameters, $\lambda_0=d$, the Bragg's law $N\lambda_0=2d\sin(\theta)$ gives the Bragg angle $\theta_\text{B}=\arcsin(1/2)=\pi/6$.  Simulations with two different laser incidence angles are checked as shown in Fig.\,\ref{fig-sketch}.

In the first case, we set $\theta_\text{B}$ as the angle between the crystal axis and the $\bm x$ axis. From the theoretical analysis, the scattered laser will be along the direction of $2\theta_B$, as shown in Fig.\,\ref{fig-sketch}-(a). In the second simulation, as a comparison, the crystal axis is along the $\bm x$ axis, as shown in Fig.\,\ref{fig-sketch}-(b). The distributions of the scattered laser fields for the two cases from the simulations are shown in Fig.\,\ref{fig-weak}(a,b). It is illustrated clearly that the scattered laser has an intense component along the Bragg angle in the first case. The corresponding wave vectors of the fields are shown in Fig.\,\ref{fig-weak}(c,d), which are consistent with the theoretical predictions. Therefore, the modified code correctly simulates the Bragg scattering, which proves the contribution of combined electrons is properly modeled.
	
\begin{figure}[htbp]
	\centering
	\includegraphics[scale=0.8]{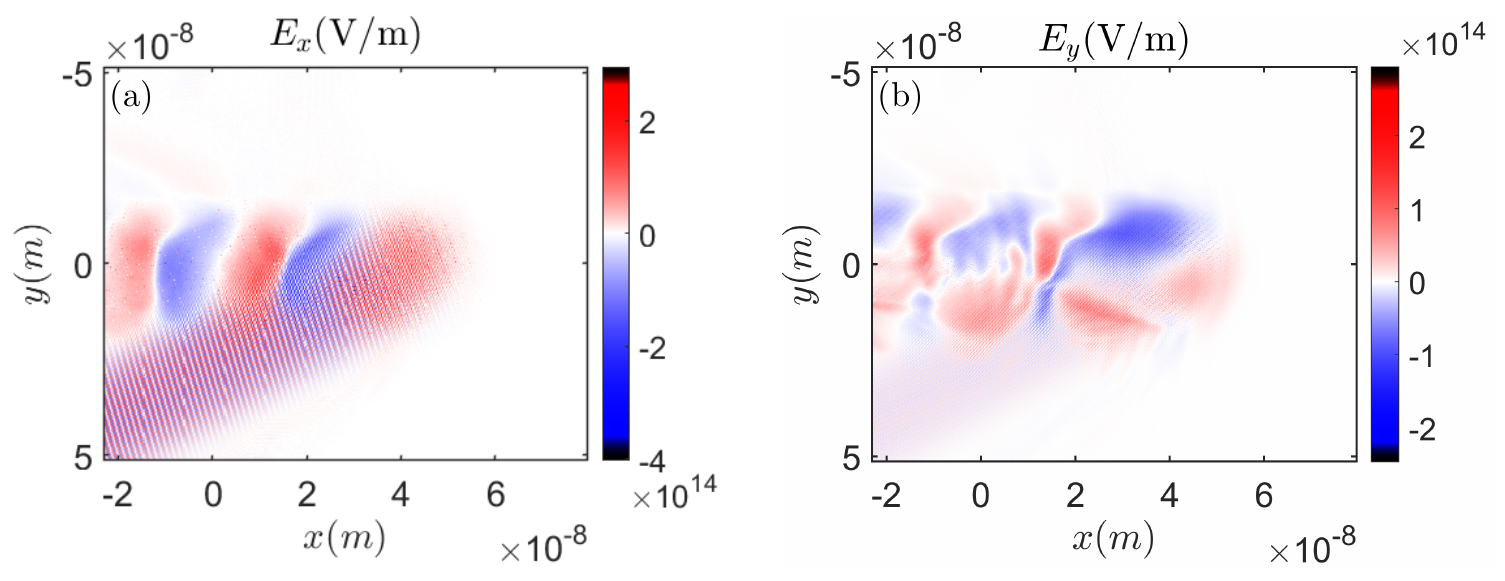}
	\caption{The electric field $E_x$ (a) and $E_y$ (b) for the simulation of laser wakefield acceleation in a crystal.\label{fig-wakefield}}
\end{figure}

\textcolor{\revisedcolor}{
As a typical application, here we show such a scattering effect on the generation of the wakefield driven by an x-ray laser in a crystal.  The parameters are the same as those in the simulation of the Bragg scattering, except the normalized vector potential $a_0$ of the laser is 3.0 and the gold is longer and pre-ionized with 30 free electrons per ion. The excited longitudinal and transverse wakefields, and the scattered laser fields are shown in Fig.\,\ref{fig-wakefield}. We can find that the laser is continuously scattered in the Bragg angle and the wakefield is slightly modulated by the scattered beam. These effects cannot be observed when the plasma is initially fully ionized and thus the response of bound electrons is absent. Detailed studies of x-ray driven wakefield acceleration in crystals, such as wake excitation and electron injection, will be our future work. }

\section{Discussion \& Summary}
\label{sec-conclusion}

\textcolor{\revisedcolor}{To model the atomic physics in PIC simulations is a quite challenging topic. It would need to solve a time-dependent Schrodinger or Dirac equation for every ion to obtain an overall model, which is impossible. However, the expectation value $\left< r\right>$ and $\left< p\right>$ from Schrodinger's or Dirac's equation is just the same as that from the Newtonian or relativistic equation of motion. In other words, $\left<r\right>$ and $\left<p\right>$ are calculated precisely in our semi-classical model. To preserve more quantum information, more properties should be calculated in our model. For example, the evolving equations of $\left<r^2\right>$ can be derived from the quantum equations and included in our model. In this way, we can preserve new information $\sigma^2(r)$ in the model. In general, we can calculate more properties and consume more computing resources to preserve more quantum information. }

\textcolor{\revisedcolor}{
As we mentioned before, although our method can be extended to include more bound electron effects, so far it is still limited for applications. The quantum effects and detailed response of each bound electron are not included. Besides these, the artificial coherence due to the PIC macro particle model should also be well treated. To fully or even partially include the quantum mechanics in a routine PIC code is extremely difficult due to both considerations of the huge number of particles and the details of each particle. This means that our model can only be used for those processes in which the quantum effects are not so obvious, i.e. the material’s response can be described in a classical way.}

In summary, by treating the partially ionized ion particles as a two-body system with an internal interaction rather than a single charged particle, the combined electron effects due to both free electrons and bound electrons can be included in a normal PIC frame. The combined electrons can reserve the dipole contribution of the electrons and thus the material's permittivity can be included without modifying the Maxwell equations.  Such response and other combined electron effects can be adjusted by modulating the internal interaction of the two-body system through the parameters $Z_b$ and $a_c$. The permittivity of a medium made by such a model is derived and compared with the experimental data. The laser propagation and scattering in neutral medium and crystals are successfully simulated with a PIC code modified with our model. This method shows a simple way to simulate the effect of the bound electrons in the PIC codes and has the potential to demonstrate and study the linear and nonlinear interactions of lasers with crystals and partially ionized medium at various wavelengths from x-rays to the mid-infrared regime. It deserves to point out that such a modified code should be used carefully since the macro-particle model for combined electrons may artificially introduce the coherence of the different electrons. Our model may be extended to high order contributions of the combined electrons by including multiple components, even though the complexity will be increased.

\section*{Acknowledgments}
\label{sec-acknowledgement}

The computations in this paper were run on the $\pi$ 2.0 cluster supported by the Center for High Performance Computing at Shanghai Jiao Tong University. This work was supported by the National Natural Science Foundation of China (11991074 and 12135009).

\bibliographystyle{elsarticle-num}
\bibliography{article_program_crystal}

\end{document}